\documentclass[aps,prl,reprint,twocolumn,superscriptaddress,raggedbottom]{revtex4-2}
\usepackage{amsmath,amssymb}
\usepackage{graphicx}
\usepackage{bm}
\usepackage{physics}
\usepackage{color}
\usepackage{tikz}
\usepackage{graphicx}
\usetikzlibrary{decorations.markings}
\usepackage{mathtools}

\NewDocumentCommand{\GM}{o}{%
  \mathrm{GM}\IfValueT{#1}{^{(#1)}}%
}

\begin{document}

\title{Multipartite Entanglement Can Probe Wormhole Moduli\\
Invisible to Bipartite Entanglement}

\author{Takanori Anegawa}
\affiliation{Yonago College, National Institute of Technology,
Yonago, Tottori 683-8502, Japan}

\author{Norihiro Iizuka}
\affiliation{Department of Physics, National Tsing Hua University, Hsinchu 300044, Taiwan}
\affiliation{Yukawa Institute for Theoretical Physics, Kyoto University, Kyoto 606-8502, Japan}

\date{\today}

\begin{abstract}
In a companion paper, we found that near the untwisted point of a four-boundary AdS$_3$ wormhole, whenever the complete bipartite Ryu--Takayanagi entropy vector is insensitive to a Fenchel--Nielsen twist, the holographic $\mathtt q=4$ multi-entropy is insensitive as well. Here we show that the situation changes at finite twist. The twist lengthens the crossing geodesics and thereby enlarges the region in which all bipartite RT entropies remain invariant. In part of this enlarged region, a twist-sensitive $\mathtt q=4$ network becomes globally minimal. We thus find an open region of wormhole moduli space in which the complete bipartite RT entropy vector is exactly unchanged while the $\mathtt q=4$ multi-entropy detects the twist. Our result provides an explicit example in which the $\mathtt q=4$ genuine multi-entropy probes bulk geometric information invisible to all bipartite entanglement entropies.
\end{abstract}

\maketitle

\section{I. Introduction}

In a companion paper (Paper I)~\cite{Iizuka:2026PartI}, we asked whether the holographic $\mathtt q=4$ multi-entropy $S^{(4)}$~\cite{Gadde:2022cqi,Penington:2022dhr,Gadde:2023zzj} can detect a bulk modulus that is exactly invisible to all bipartite Ryu--Takayanagi (RT) entropies~\cite{Ryu:2006bv,Ryu:2006ef}. 
For an explicit family of four-boundary AdS$_3$ wormholes related by a Fenchel--Nielsen twist $\tau$~\cite{Wolpert:1982},  Paper I showed that, along the symmetric slice $\ell=m$ of parameter space and for sufficiently small $|\tau|$, the answer is negative: whenever the complete bipartite RT entropy vector is exactly $\tau$-independent, the globally minimal $\mathtt q=4$ network is itself $\tau$-independent. Thus $S^{(4)}$ fails to detect the twist as well. 
That analysis was explicitly restricted to a neighborhood of $\tau=0$; it remained open whether a finite twist could change this conclusion.

\begin{figure}[t]
\centering
\includegraphics[width=0.49\textwidth]{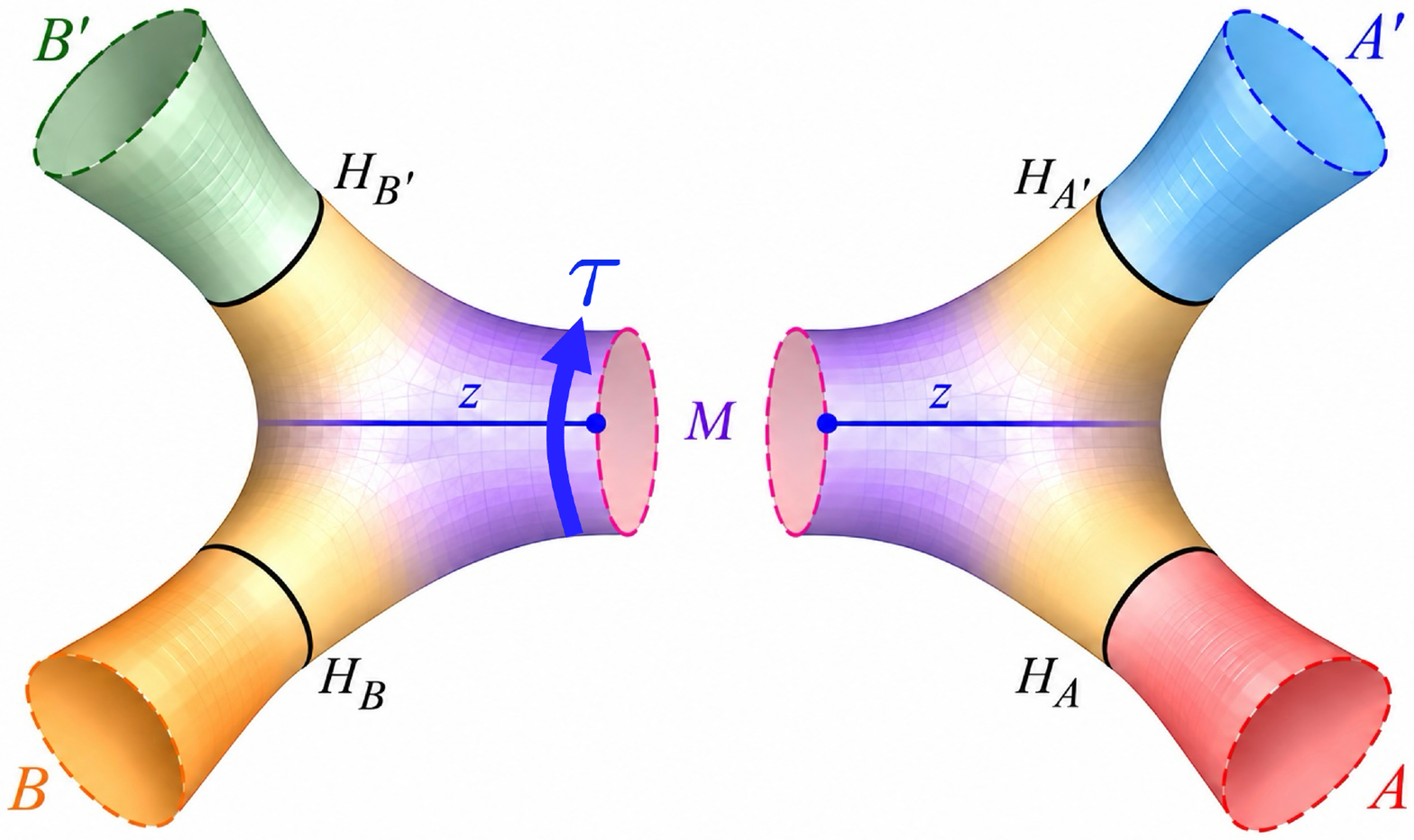}
\caption{
Geometric origin of the twist dependence of the crossing geodesic.
At zero twist, the two minimizing segments of length $z$ meet on the internal cuff $M$ and form a closed crossing geodesic of length $2z$. For $0<\tau\le m/4$, the Fenchel–Nielsen twist displaces their endpoints along $M$, so the closed geodesic must readjust and cross the two pairs of pants obliquely. The increasing endpoint mismatch makes the crossing length twist dependent and larger than $2z$. (Figure prepared with the assistance of ChatGPT.)
}
\label{fig:four-boundary-wormhole}
\end{figure}

This is not merely a technical extension. In Paper I, the twist-blindness of bipartite RT followed from a lower bound, $2z$, on the length of any connected curve realizing a crossing bipartition; this bound is saturated only at $\tau=0$; for $\tau\neq0$ the actual shortest such curve is longer than $2z$, since the twist misaligns the endpoints at which the two halves of the curve would otherwise join smoothly across the internal cuff $M$. See Fig.~\ref{fig:four-boundary-wormhole}. 
Since the disconnected competitor is unchanged, the bipartite twist-invariant window can widen away from $\tau=0$.

In this paper we show that this is exactly what happens. Along the symmetric slice $\ell=m$, we track $\tau$ over its full fundamental domain and find that the bipartite-invariant window in $\ell$, which
Paper~I identified as $\ell\lesssim\ell_*\simeq3.671274$ near $\tau=0$, widens to $\ell\lesssim\ell_c\simeq4.018428$ at the symmetric point $\tau=m/4$. 
Within $3.9385<\ell<\ell_c$, the globally minimal $\mathtt q=4$ network at $\tau=m/4$ is a twist-sensitive mixed-channel configuration, while the complete bipartite RT entropy vector is exactly twist-invariant; an explicit example is $\ell=m=4,\ \tau=1$. 
In the rest of the strip $\ell_*<\ell<\ell_c$, a $\tau$-independent hybrid configuration  remains minimal. 
We further find that this phenomenon is not confined
to a single point or to the symmetric slice $\ell=m$: it persists
over an open region of $(\ell,m)$ parameter space at fixed
$\tau=m/4$, and survives even in the limit $\ell\to\infty$ within a
narrow but finite window of $m$.

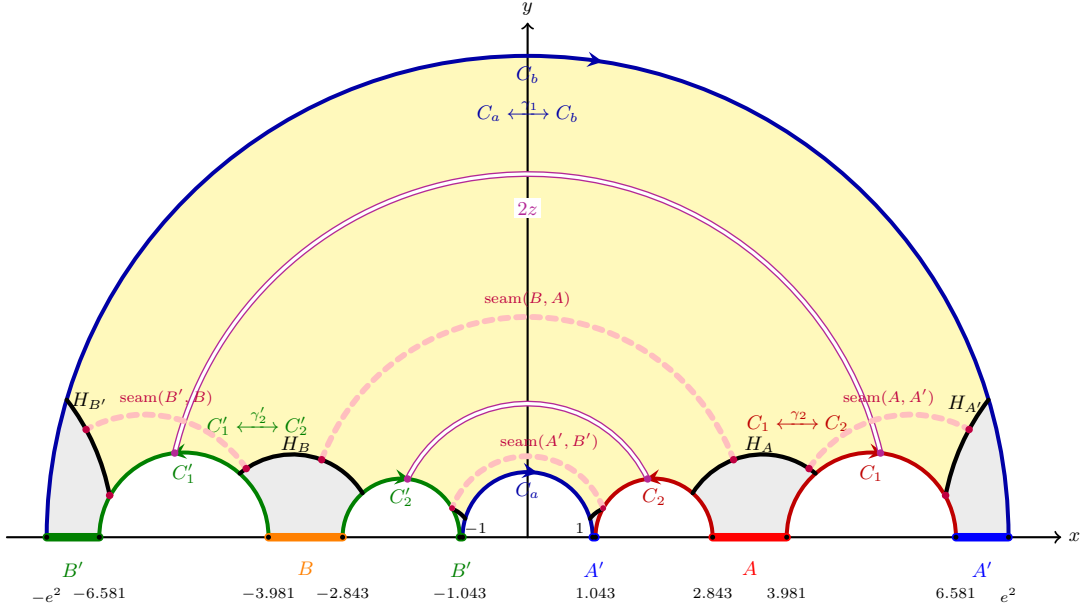
\begin{figure*}[t]
\centering

\begin{tikzpicture}[
    x=1.05cm,
    y=1.05cm,
    scale=0.82,
    transform shape,
    line cap=round,
    line join=round,
    arrow along path/.style={
        postaction={
            decorate,
            decoration={
                markings,
                mark=at position 0.55 with
                {\arrow{stealth}}
            }
        }
    }
]

\def\muR{7.389056}   
\def\cOne{5.28094}
\def\rOne{1.3}
\def\cTwo{1.94275}
\def\rTwo{0.9}

\fill[yellow!33]
    (-\muR,0)
    arc[start angle=180,end angle=0,radius=\muR]
    -- cycle;

\fill[white]
    (-1,0)
    arc[start angle=180,end angle=0,radius=1]
    -- cycle;

\fill[white]
    ({\cOne-\rOne},0)
    arc[start angle=180,end angle=0,radius=\rOne]
    -- cycle;

\fill[white]
    ({\cTwo-\rTwo},0)
    arc[start angle=180,end angle=0,radius=\rTwo]
    -- cycle;

\fill[white]
    ({-\cTwo-\rTwo},0)
    arc[start angle=180,end angle=0,radius=\rTwo]
    -- cycle;

\fill[white]
    ({-\cOne-\rOne},0)
    arc[start angle=180,end angle=0,radius=\rOne]
    -- cycle;

\fill[gray!15]
    (2.84275,0)
    -- (3.98094,0)
    arc[start angle=180,end angle=131.2265,radius=1.3]
    arc[start angle=50.2790,end angle=147.8256,radius=1.2711745]
    arc[start angle=48.7734,end angle=0,radius=0.9]
    -- cycle;

\fill[gray!15]
    (-3.98094,0)
    -- (-2.84275,0)
    arc[start angle=180,end angle=131.2266,radius=0.9]
    arc[start angle=32.1744,end angle=129.7210,radius=1.2711745]
    arc[start angle=48.7735,end angle=0,radius=1.3]
    -- cycle;

\fill[gray!15]
    (6.58094,0)
    -- (7.389056,0)
    arc[start angle=0,end angle=16.5902,radius=7.389056]
    arc[start angle=142.3695,end angle=169.5508,radius=3.4554105]
    arc[start angle=28.8205,end angle=0,radius=1.3]
    -- cycle;

\fill[gray!15]
    (1,0)
    -- (1.04275,0)
    arc[start angle=180,end angle=151.1792,radius=0.9]
    arc[start angle=111.9092,end angle=142.3695,radius=0.4676390]
    arc[start angle=16.5902,end angle=0,radius=1]
    -- cycle;

\fill[gray!15]
    (-7.389056,0)
    -- (-6.58094,0)
    arc[start angle=180,end angle=151.1795,radius=1.3]
    arc[start angle=10.4492,end angle=37.6305,radius=3.4554105]
    arc[start angle=163.4098,end angle=180,radius=7.389056]
    -- cycle;

\fill[gray!15]
    (-1.04275,0)
    -- (-1,0)
    arc[start angle=180,end angle=163.4098,radius=1]
    arc[start angle=37.6305,end angle=68.0908,radius=0.4676390]
    arc[start angle=28.8208,end angle=0,radius=0.9]
    -- cycle;

\draw[->,line width=0.8pt]
    (-8.0,0) -- (8.2,0)
    node[right] {$x$};

\draw[->,line width=0.8pt]
    (0,0) -- (0,7.9)
    node[above] {$y$};

\draw[blue!65!black,line width=1.5pt,arrow along path]
    (-\muR,0)
    arc[start angle=180,end angle=0,radius=\muR];

\draw[blue!65!black,line width=1.5pt,arrow along path]
    (-1,0)
    arc[start angle=180,end angle=0,radius=1];

\draw[red!75!black,line width=1.5pt,arrow along path]
    ({\cOne-\rOne},0)
    arc[start angle=180,end angle=0,radius=\rOne];

\draw[red!75!black,line width=1.5pt,arrow along path]
    ({\cTwo+\rTwo},0)
    arc[start angle=0,end angle=180,radius=\rTwo];

\draw[green!50!black,line width=1.5pt,arrow along path]
    ({-\cOne+\rOne},0)
    arc[start angle=0,end angle=180,radius=\rOne];

\draw[green!50!black,line width=1.5pt,arrow along path]
    ({-\cTwo-\rTwo},0)
    arc[start angle=180,end angle=0,radius=\rTwo];

\node[blue!65!black,font=\small] at (0,7.10) {$C_b$};
\node[blue!65!black,font=\small] at (0,0.72) {$C_a$};

\node[red!75!black,font=\small]
    at (\cOne,0.98) {$C_1$};

\node[red!75!black,font=\small]
    at (\cTwo,0.62) {$C_2$};

\node[green!50!black,font=\small]
    at (-\cTwo,0.62) {$C_2'$};

\node[green!50!black,font=\small]
    at (-\cOne,0.98) {$C_1'$};

\node[blue!65!black,font=\small]
    at (0,6.55)
    {$C_a \xleftrightarrow{\ \gamma_1\ } C_b$};

\node[red!75!black,font=\small]
    at (4.15,1.80)
    {$C_1 \xleftrightarrow{\ \gamma_2\ } C_2$};

\node[green!50!black,font=\small]
    at (-4.15,1.80)
    {$C_1' \xleftrightarrow{\ \gamma_2'\ } C_2'$};

\draw[green!50!black,line width=3.2pt]
    (-7.389056,0) -- (-6.58094,0);

\draw[green!50!black,line width=3.2pt]
    (-1.04275,0) -- (-1,0);

\draw[orange,line width=3.2pt]
    (-3.98094,0) -- (-2.84275,0);

\draw[blue,line width=3.2pt]
    (1,0) -- (1.04275,0);

\draw[blue,line width=3.2pt]
    (6.58094,0) -- (7.389056,0);

\draw[red,line width=3.2pt]
    (2.84275,0) -- (3.98094,0);

\node[green!50!black,font=\small,below=8pt]
    at ({(-7.389056-6.58094)/2},0) {$B'$};

\node[green!50!black,font=\small,below=8pt]
    at ({(-1.04275-1)/2},0) {$B'$};

\node[orange,font=\small,below=8pt]
    at ({(-3.98094-2.84275)/2},0) {$B$};

\node[blue,font=\small,below=8pt]
    at ({(1+1.04275)/2},0) {$A'$};

\node[blue,font=\small,below=8pt]
    at ({(6.58094+7.389056)/2},0) {$A'$};

\node[red,font=\small,below=8pt]
    at ({(2.84275+3.98094)/2},0) {$A$};

\foreach \x in {
    -7.389056,-6.58094,-3.98094,-2.84275,
    -1.04275,-1,
     1,1.04275,2.84275,3.98094,6.58094,7.389056
}{
    \fill (\x,0) circle (1.15pt);
}

\node[below=20pt,font=\scriptsize]
    at (-7.389056,0) {$-e^2$};

\node[below=20pt,font=\scriptsize]
    at (-6.58094,0) {$-6.581$};

\node[below=20pt,font=\scriptsize]
    at (-3.98094,0) {$-3.981$};

\node[below=20pt,font=\scriptsize]
    at (-2.84275,0) {$-2.843$};

\node[below=20pt,font=\scriptsize]
    at (-1.04275,0) {$-1.043$};

\node[below=20pt,font=\scriptsize]
    at (-0.8,1) {$-1$};

\node[below=20pt,font=\scriptsize]
    at (0.8,1) {$1$};

\node[below=20pt,font=\scriptsize]
    at (1.04275,0) {$1.043$};
    
\node[below=20pt,font=\scriptsize]
    at (2.84275,0) {$2.843$};

\node[below=20pt,font=\scriptsize]
    at (3.98094,0) {$3.981$};

\node[below=20pt,font=\scriptsize]
    at (6.58094,0) {$6.581$};

\node[below=20pt,font=\scriptsize]
    at (7.389056,0) {$e^2$};

\draw[black,line width=1.6pt]
    (2.535885,0.676898)
    arc[
        start angle=147.8256,
        end angle=50.2790,
        radius=1.2711745
    ];

\node[black,font=\small]
    at (3.56,1.43) {$H_A$};

\draw[black,line width=1.6pt]
    (6.419914,0.626687)
    arc[
        start angle=169.5508,
        end angle=142.3695,
        radius=3.4554105
    ];

\node[black,font=\small]
    at (6.73,2.05) {$H_{A'}$};

\draw[black,line width=1.6pt]
    (-4.424191,0.977743)
    arc[
        start angle=129.7210,
        end angle=32.1744,
        radius=1.2711745
    ];

\node[black,font=\small]
    at (-3.56,1.43) {$H_B$};

\draw[black,line width=1.6pt]
    (-7.081459,2.109761)
    arc[
        start angle=37.6305,
        end angle=10.4492,
        radius=3.4554105
    ];

\node[black,font=\small]
    at (-6.73,2.1) {$H_{B'}$};

\draw[black,line width=1.6pt]
    (1.154231,0.433864)
    arc[
        start angle=111.9092,
        end angle=142.3695,
        radius=0.4676390
    ];

\draw[black,line width=1.6pt]
    (-0.958371,0.285525)
    arc[
        start angle=37.6305,
        end angle=68.0908,
        radius=0.4676390
    ];


%

\definecolor{zcrosscolor}{RGB}{180,40,150}

\tikzset{
    zcross/.style={
        zcrosscolor,
        double,
        double distance=1.2pt,
        line width=0.5pt
    }
}

\def\rhoInner{2.050535704}
\def\rhoOuter{5.573931808}

\def\xInner{1.845058354}
\def\yInner{0.894682258}

\def\xOuter{5.422050155}
\def\yOuter{1.292318817}


\pgfmathsetmacro{\thetaInner}
    {atan2(\yInner,\xInner)}

\draw[zcross]
    (\xInner,\yInner)
    arc[
        start angle=\thetaInner,
        end angle={180-\thetaInner},
        radius=\rhoInner
    ];


\pgfmathsetmacro{\thetaOuter}
    {atan2(\yOuter,\xOuter)}

\draw[zcross]
    (\xOuter,\yOuter)
    arc[
        start angle=\thetaOuter,
        end angle={180-\thetaOuter},
        radius=\rhoOuter
    ];


\fill[zcrosscolor]
    (\xInner,\yInner) circle (1.7pt);

\fill[zcrosscolor]
    (-\xInner,\yInner) circle (1.7pt);

\fill[zcrosscolor]
    (\xOuter,\yOuter) circle (1.7pt);

\fill[zcrosscolor]
    (-\xOuter,\yOuter) circle (1.7pt);


\node[
    zcrosscolor,
    font=\small,
    fill=white,
    inner sep=1.5pt
]
    at (0,5.05) {$2z$};



\draw[pink,dashed,line width=2.2pt]
    (4.32326,1.05345)
    arc[
        start angle=145.968,
        end angle=61.421,
        radius=1.88231
    ];

\draw[pink,dashed,line width=2.2pt]
    (-4.32326,1.05345)
    arc[
        start angle=34.032,
        end angle=118.579,
        radius=1.88231
    ];

\draw[pink,dashed,line width=2.2pt]
    (-3.16447,1.18982)
    arc[
        start angle=159.394,
        end angle=20.606,
        radius=3.38076
    ];

\draw[pink,dashed,line width=2.2pt]
    (1.161009,0.445961)
    arc[
        start angle=21.0125,
        end angle=158.9875,
        radius=1.243713
    ];

\draw[pink,dashed,line width=2.2pt]
    (6.422752,0.641906)
    arc[
        start angle=79.2942,
        end angle=80.4196,
        radius=0.653277
    ];

\draw[pink,dashed,line width=2.2pt]
    (-6.410121,0.644166)
    arc[
        start angle=99.5804,
        end angle=100.7058,
        radius=0.653277
    ];

\fill[purple] (4.32326,1.05345) circle (1.7pt);
\fill[purple] (6.78362,1.65296) circle (1.7pt);

\fill[purple] (-6.78362,1.65296) circle (1.7pt);
\fill[purple] (-4.32326,1.05345) circle (1.7pt);

\fill[purple] (-3.16447,1.18982) circle (1.7pt);
\fill[purple] (3.16447,1.18982) circle (1.7pt);

\fill[purple] (6.422752,0.641906) circle (1.7pt);
\fill[purple] (-6.422752,0.641906) circle (1.7pt);

\fill[purple]
    (1.161009,0.445961) circle (1.35pt);

\fill[purple]
    (-1.161009,0.445961) circle (1.35pt);

\fill[purple]
    (6.410121,0.644166) circle (1.35pt);

\fill[purple]
    (-6.410121,0.644166) circle (1.35pt);

\node[purple,font=\scriptsize]
    at (5.55,2.15)
    {$\mathrm{seam}(A,A')$};

\node[purple,font=\scriptsize]
    at (-5.55,2.15)
    {$\mathrm{seam}(B',B)$};

\node[purple,font=\scriptsize]
    at (0,3.65)
    {$\mathrm{seam}(B,A)$};

\node[purple,font=\scriptsize,xshift=10pt]
    at (0,1.47)
    {$\mathrm{seam}(A',B')$};

\end{tikzpicture}

\caption{
The closed crossing geodesic of length $2z$ at $\tau=0$ for
$\ell=m=2$, shown by the double magenta curves. The inner arc is the invariant axis of $\Gamma_N=\gamma_2(\gamma_2')^{-1}$, while the outer arc is its image under $\gamma_2^{-1}$. The two arcs are identified across the paired circles and therefore form a single closed geodesic in the quotient geometry. For the parameters shown here, the inner and outer arcs are centered at the origin with radii $2.05054$ and $5.57393$, respectively, and their total hyperbolic length is $2z=7.22445$. (Figure prepared
with the assistance of Claude and ChatGPT.)
}
\label{fig:schottky-fundamental-region}

\end{figure*}

\section{II. Setup}
We use the same four-boundary AdS$_3$ wormhole and Schottky-group
construction as in Paper I~\cite{Iizuka:2026PartI}, which we briefly recall.
Such multiboundary AdS$_3$ wormholes provide a useful setting for
studying the relation between entanglement and bulk geometry and
moduli~\cite{Balasubramanian:2014hda,Caceres:2019giy, Anegawa:2020lzw}.
The time-symmetric
slice is a hyperbolic surface $\Sigma=\mathbb H^2/\Gamma$,
$\Gamma=\langle\gamma_1,\gamma_2,\gamma_2'\rangle\subset SL(2,\mathbb
R)$, with
\begin{equation}
\gamma_1=\begin{pmatrix}\mu&0\\0&\mu^{-1}\end{pmatrix},
\quad
\gamma_2=\frac{1}{\sqrt{R_1R_2}}\begin{pmatrix}-c_2&c_1c_2+R_1R_2\\-1&c_1\end{pmatrix},
\end{equation}
and $\gamma_2'$ defined analogously from $c_1'=-c_1,\,c_2'=-c_2,\,R_1'=R_1,\,R_2'=R_2$.
The four asymptotic horizons $A,A',B,B'$ and the internal cuff $M$
have lengths $L_A,L_{A'},L_B,L_{B'}$ and $L_M=2\log\mu$ as in
Paper~I, Sec.~II. The Fenchel--Nielsen twist is implemented by
conjugating the second pair of pants,
\begin{equation}
\gamma_2'(\tau)=\eta(\tau)\,\gamma_2'\,\eta(-\tau),
\qquad
\eta(\tau)=\begin{pmatrix}e^{\tau/2}&0\\0&e^{-\tau/2}\end{pmatrix},
\end{equation}
which leaves $L_A$, $L_{A'}$, $L_B$, $L_{B'}$, $L_M$ exactly $\tau$-independent.
As in Paper~I, we first consider the symmetric family
$L_A=L_{A'}=L_B=L_{B'}=\ell$, $L_M=m$. We begin with the
one-parameter slice $\ell=m$, before relaxing this restriction in
Sec.~VI.

As is clear from Fig.~\ref{fig:four-boundary-wormhole}, reversing
the direction of the Fenchel--Nielsen twist corresponds to
reflecting the geometry. More explicitly, the reflection
$z\mapsto-\bar z$, followed by the dilation
$z\mapsto e^{-\tau}z$, maps the group at $\tau$ to the group at
$-\tau$, under which
\begin{equation}
\gamma_1\longmapsto\gamma_1,\quad
\gamma_2\longmapsto\gamma_2'(-\tau),\quad
\gamma_2'(\tau)\longmapsto\gamma_2,
\end{equation} 
hence
$A\leftrightarrow B$ and $A'\leftrightarrow B'$. This exchange
maps each of the crossing partitions $AB|A'B'$ and $AB'|A'B$ to
itself, and $L_A=L_B$, $L_{A'}=L_{B'}$ in the symmetric family.
Since a reflection is an isometry, the relevant geodesic lengths are even functions of $\tau$, $L(\tau)=L(-\tau)$. 
Paper~I also established a model-independent lower bound: any simple
closed curve $\Gamma$ realizing a crossing bipartition ($AB|A'B'$ or
$AB'|A'B$) satisfies $L(\Gamma)\geq2z$, where $z=z(\ell,m)$ is the
$\tau$-independent length of the shortest returning orthogeodesic in
either pair of pants. 
At $\tau=0$, the two such arcs join across $M$
to form the closed crossing geodesic of total length $2z$. In the Schottky parametrization above,
\begin{equation}
z=
\operatorname{arccosh}\!\left(\frac{r+s}{s-r}\right),
\quad
r=c_2,\quad
s=c_2+\frac{R_1R_2}{c_1}.
\label{eq:zformula}
\end{equation}
Whenever
\begin{equation}
2z(\ell,m)>2\ell,
\label{eq:crossing-condition}
\end{equation}
no such curve can compete with the disconnected pair of horizons,
and the complete bipartite RT entropy vector is exactly
$\tau$-independent,
\begin{align}
\label{eq:SAetc}
& S(A)= S(A')= S(B)= S(B')  = \frac{\ell}{4G_N},
\\
& S(AA')= \frac{\min\{2\ell,m\}}{4G_N},
\quad  S(AB)= S(AB')= \frac{2\ell}{4G_N}.
\label{eq:SABetc}
\end{align}
Paper~I used this $\tau$-independent sufficient condition to
identify the boundary $\ell_*\simeq3.671274$ of the guaranteed
bipartite-invariant region on the symmetric slice. As we show below,
at finite twist the actual shortest crossing geodesic can be
strictly longer than the lower bound $2z$, thereby extending this
region beyond $\ell_*$.

The effect of the twist on the corresponding horizon geodesics is
particularly simple. We first recall that conjugation maps the
invariant axis of a hyperbolic element in the same way: if
$g'=hgh^{-1}$, then
\begin{equation}
\operatorname{Axis}(g')
=
h\bigl(\operatorname{Axis}(g)\bigr).
\end{equation}
Indeed, denoting $\mathcal A=\operatorname{Axis}(g)$, so that
$g(\mathcal A)=\mathcal A$, we have
\begin{equation}
g'\bigl(h(\mathcal A)\bigr)
=
hgh^{-1}\bigl(h(\mathcal A)\bigr)
=
hg(\mathcal A)
=
h(\mathcal A).
\end{equation}
Applying this observation to $g(\tau)=\eta(\tau)g\eta(-\tau)$, we obtain
\begin{equation}
\operatorname{Axis}(g(\tau))
=
\eta(\tau)\operatorname{Axis}(g).
\end{equation}
Since $\eta(\tau)$ acts on the upper half-plane as
\begin{equation}
z\longmapsto e^\tau z,
\end{equation}
the corresponding horizon lifts are therefore mapped by this simple
rescaling. In particular, $H_B$ and $H_{B'}$ are mapped by
$z\mapsto e^\tau z$, while $H_A$ and $H_{A'}$ remain unchanged.


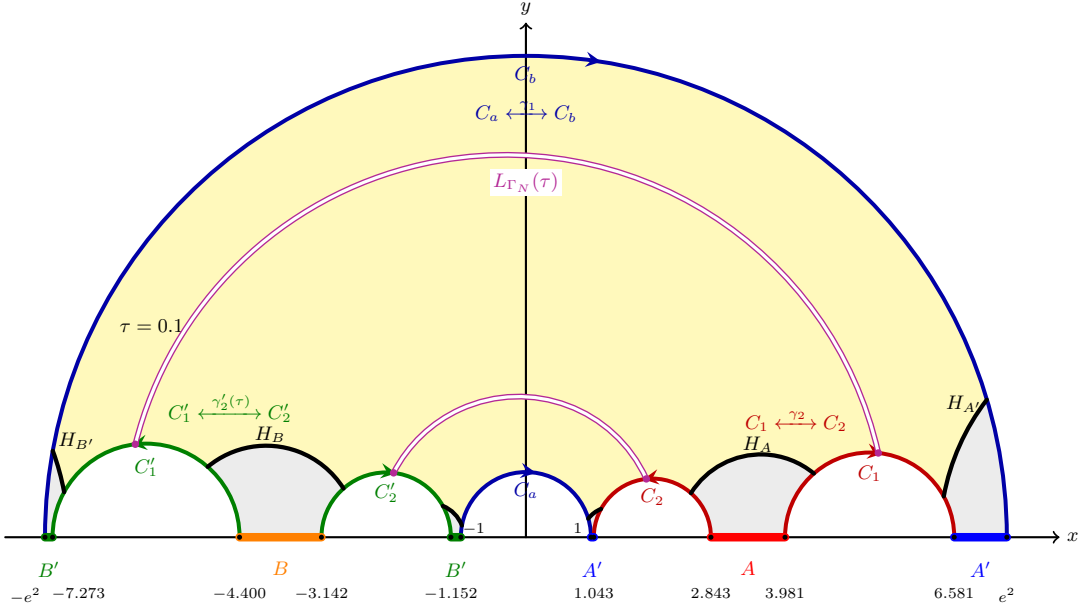
\begin{figure*}[t]
\centering

\begin{tikzpicture}[
    x=1.05cm,
    y=1.05cm,
    scale=0.82,
    transform shape,
    line cap=round,
    line join=round,
    arrow along path/.style={
        postaction={
            decorate,
            decoration={
                markings,
                mark=at position 0.55 with
                {\arrow{stealth}}
            }
        }
    }
]


\def\muR{7.389056}   
\def\cOne{5.28094}
\def\rOne{1.3}
\def\cTwo{1.94275}
\def\rTwo{0.9}

\def\tauTwist{0.1}
\pgfmathsetmacro{\twistScale}{exp(\tauTwist)}

\pgfmathsetmacro{\cOneP}{\twistScale*\cOne}
\pgfmathsetmacro{\rOneP}{\twistScale*\rOne}
\pgfmathsetmacro{\cTwoP}{\twistScale*\cTwo}
\pgfmathsetmacro{\rTwoP}{\twistScale*\rTwo}


\fill[yellow!33]
    (-\muR,0)
    arc[start angle=180,end angle=0,radius=\muR]
    -- cycle;

\fill[white]
    (-1,0)
    arc[start angle=180,end angle=0,radius=1]
    -- cycle;

\fill[white]
    ({\cOne-\rOne},0)
    arc[start angle=180,end angle=0,radius=\rOne]
    -- cycle;

\fill[white]
    ({\cTwo-\rTwo},0)
    arc[start angle=180,end angle=0,radius=\rTwo]
    -- cycle;

\fill[white]
    ({-\cTwoP-\rTwoP},0)
    arc[start angle=180,end angle=0,radius=\rTwoP]
    -- cycle;

\fill[white]
    ({-\cOneP-\rOneP},0)
    arc[start angle=180,end angle=0,radius=\rOneP]
    -- cycle;


\fill[gray!15]
    (2.84275,0)
    -- (3.98094,0)
    arc[start angle=180,end angle=131.2265,radius=1.3]
    arc[start angle=50.2790,end angle=147.8256,radius=1.2711745]
    arc[start angle=48.7734,end angle=0,radius=0.9]
    -- cycle;


\fill[gray!15]
    (-4.399619115,0)
    -- (-3.141724627,0)
    arc[
        start angle=180,
        end angle=131.2266,
        radius=0.994653826
    ]
    arc[
        start angle=32.1744,
        end angle=129.7210,
        radius=1.404865089
    ]
    arc[
        start angle=48.7735,
        end angle=0,
        radius=1.436722193
    ]
    -- cycle;


\fill[gray!15]
    (6.58094,0)
    -- (7.389056,0)
    arc[start angle=0,end angle=16.5902,radius=7.389056]
    arc[start angle=142.3695,end angle=169.5508,radius=3.4554105]
    arc[start angle=28.8205,end angle=0,radius=1.3]
    -- cycle;


\fill[gray!15]
    (1,0)
    -- (1.04275,0)
    arc[start angle=180,end angle=151.1792,radius=0.9]
    arc[start angle=111.9092,end angle=142.3695,radius=0.4676390]
    arc[start angle=16.5902,end angle=0,radius=1]
    -- cycle;

%
%

\fill[gray!15]
    (-7.389056,0)
    -- (-7.273063502,0)
    arc[
        start angle=180,
        end angle=151.1795,
        radius=1.436722193
    ]
    arc[
        start angle=10.4492,
        end angle=20.3099,
        radius=3.818819195
    ]
    arc[
        start angle=169.6659,
        end angle=180,
        radius=7.389056
    ]
    -- cycle;

%
%

\fill[gray!15]
    (-1.152416975,0)
    -- (-1,0)
    arc[
        start angle=180,
        end angle=169.6659,
        radius=1
    ]
    arc[
        start angle=20.3100,
        end angle=68.0908,
        radius=0.516821023
    ]
    arc[
        start angle=28.8208,
        end angle=0,
        radius=0.994653826
    ]
    -- cycle;


\draw[->,line width=0.8pt]
    (-8.0,0) -- (8.2,0)
    node[right] {$x$};

\draw[->,line width=0.8pt]
    (0,0) -- (0,7.9)
    node[above] {$y$};


\draw[blue!65!black,line width=1.5pt,arrow along path]
    (-\muR,0)
    arc[start angle=180,end angle=0,radius=\muR];

\draw[blue!65!black,line width=1.5pt,arrow along path]
    (-1,0)
    arc[start angle=180,end angle=0,radius=1];

\draw[red!75!black,line width=1.5pt,arrow along path]
    ({\cOne-\rOne},0)
    arc[start angle=180,end angle=0,radius=\rOne];

\draw[red!75!black,line width=1.5pt,arrow along path]
    ({\cTwo+\rTwo},0)
    arc[start angle=0,end angle=180,radius=\rTwo];

\draw[green!50!black,line width=1.5pt,arrow along path]
    ({-\cOneP+\rOneP},0)
    arc[start angle=0,end angle=180,radius=\rOneP];

\draw[green!50!black,line width=1.5pt,arrow along path]
    ({-\cTwoP-\rTwoP},0)
    arc[start angle=180,end angle=0,radius=\rTwoP];


\node[blue!65!black,font=\small]
    at (0,7.10) {$C_b$};

\node[blue!65!black,font=\small]
    at (0,0.72) {$C_a$};

\node[red!75!black,font=\small]
    at (\cOne,0.98) {$C_1$};

\node[red!75!black,font=\small]
    at (\cTwo,0.62) {$C_2$};

\node[green!50!black,font=\small]
    at (-\cTwoP,0.69) {$C_2'$};

\node[green!50!black,font=\small]
    at (-\cOneP,1.08) {$C_1'$};

\node[blue!65!black,font=\small]
    at (0,6.55)
    {$C_a \xleftrightarrow{\ \gamma_1\ } C_b$};

\node[red!75!black,font=\small]
    at (4.15,1.80)
    {$C_1 \xleftrightarrow{\ \gamma_2\ } C_2$};

\node[green!50!black,font=\small]
    at (-4.55,2.00)
    {$C_1' \xleftrightarrow{\ \gamma_2'(\tau)\ } C_2'$};


\draw[green!50!black,line width=3.2pt]
    (-7.389056,0) -- (-7.273063502,0);

\draw[green!50!black,line width=3.2pt]
    (-1.152416975,0) -- (-1,0);

\draw[orange,line width=3.2pt]
    (-4.399619115,0) -- (-3.141724627,0);

\draw[blue,line width=3.2pt]
    (1,0) -- (1.04275,0);

\draw[blue,line width=3.2pt]
    (6.58094,0) -- (7.389056,0);

\draw[red,line width=3.2pt]
    (2.84275,0) -- (3.98094,0);


\node[green!50!black,font=\small,below=8pt]
    at ({(-7.389056-7.273063502)/2},0) {$B'$};

\node[green!50!black,font=\small,below=8pt]
    at ({(-1.152416975-1)/2},0) {$B'$};

\node[orange,font=\small,below=8pt]
    at ({(-4.399619115-3.141724627)/2},0) {$B$};

\node[blue,font=\small,below=8pt]
    at ({(1+1.04275)/2},0) {$A'$};

\node[blue,font=\small,below=8pt]
    at ({(6.58094+7.389056)/2},0) {$A'$};

\node[red,font=\small,below=8pt]
    at ({(2.84275+3.98094)/2},0) {$A$};


\foreach \x in {
    -7.389056,-7.273063502,
    -4.399619115,-3.141724627,
    -1.152416975,-1,
     1,1.04275,
     2.84275,3.98094,
     6.58094,7.389056
}{
    \fill (\x,0) circle (1.15pt);
}


\node[below=20pt,font=\scriptsize]
    at (-7.689056,0) {$-e^2$};

\node[below=20pt,font=\scriptsize]
    at (-6.873063502,0) {$-7.273$};

\node[below=20pt,font=\scriptsize]
    at (-4.399619115,0) {$-4.400$};

\node[below=20pt,font=\scriptsize]
    at (-3.141724627,0) {$-3.142$};

\node[below=20pt,font=\scriptsize]
    at (-1.152416975,0) {$-1.152$};

\node[below=20pt,font=\scriptsize]
    at (-0.8,1) {$-1$};

\node[below=20pt,font=\scriptsize]
    at (0.8,1) {$1$};

\node[below=20pt,font=\scriptsize]
    at (1.04275,0) {$1.043$};

\node[below=20pt,font=\scriptsize]
    at (2.84275,0) {$2.843$};

\node[below=20pt,font=\scriptsize]
    at (3.98094,0) {$3.981$};

\node[below=20pt,font=\scriptsize]
    at (6.58094,0) {$6.581$};

\node[below=20pt,font=\scriptsize]
    at (7.389056,0) {$e^2$};


\draw[black,line width=1.6pt]
    (2.535885,0.676898)
    arc[
        start angle=147.8256,
        end angle=50.2790,
        radius=1.2711745
    ];

\node[black,font=\small]
    at (3.56,1.43) {$H_A$};


\draw[black,line width=1.6pt]
    (6.419914,0.626687)
    arc[
        start angle=169.5508,
        end angle=142.3695,
        radius=3.4554105
    ];

\node[black,font=\small]
    at (6.73,2.05) {$H_{A'}$};


\draw[black,line width=1.6pt]
    (1.154231,0.433864)
    arc[
        start angle=111.9092,
        end angle=142.3695,
        radius=0.4676390
    ];

%

\draw[black,line width=1.6pt]
    (-4.889487229,1.080573129)
    arc[
        start angle=129.7210,
        end angle=32.1744,
        radius=1.404865089
    ];

\node[black,font=\small]
    at (-3.93,1.58) {$H_B$};

%

\draw[black,line width=1.6pt]
    (-7.26919417,1.32550547)
    arc[
        start angle=20.3099,
        end angle=10.4492,
        radius=3.818819195
    ];

\node[black,font=\small]
    at (-6.9,1.48) {$H_{B'}$};

%

\draw[black,line width=1.6pt]
    (-0.98377836,0.17938822)
    arc[
        start angle=20.3100,
        end angle=68.0908,
        radius=0.516821023
    ];


\node[font=\small]
    at (-5.75,3.25) {$\tau=0.1$};




\definecolor{zcrosscolor}{RGB}{180,40,150}

\tikzset{
    zcross/.style={
        zcrosscolor,
        double,
        double distance=1.2pt,
        line width=0.5pt
    }
}

%
%
%

\draw[zcross]
    (1.850464644,0.895256060)
    arc[
        start angle=24.5020330,
        end angle=152.7589864,
        radius=2.158671207
    ];

%
%
%

\draw[zcross]
    (5.414241070,1.293147642)
    arc[
        start angle=12.7312077,
        end angle=165.9226441,
        radius=5.867874467
    ];


\fill[zcrosscolor]
    (1.850464644,0.895256060) circle (1.7pt);

\fill[zcrosscolor]
    (-2.033061491,0.988098230) circle (1.7pt);

\fill[zcrosscolor]
    (5.414241070,1.293147642) circle (1.7pt);

\fill[zcrosscolor]
    (-6.001021423,1.427252998) circle (1.7pt);


\node[
    zcrosscolor,
    font=\small,
    fill=white,
    inner sep=1.5pt
]
    at (0,5.45)
    {$L_{\Gamma_N}(\tau)$};

\end{tikzpicture}

\caption{
Schottky fundamental domain at $\ell=m=2$  after a Fenchel--Nielsen twist
$\tau=0.1$, together with the corresponding closed crossing
geodesic (magenta).
The twist is implemented by
$\gamma_2'(\tau)=\eta(\tau)\gamma_2'\eta(-\tau)$, where
$\eta(\tau)$ acts on the upper half-plane as $z\mapsto e^\tau z$.
Consequently, the primed side-pairing circles $C_1'$ and $C_2'$,
as well as the corresponding horizon axes $H_B$ and $H_{B'}$,
are mapped by the same rescaling, while the unprimed circles and
$C_a,C_b$ remain fixed.
The magenta curve is obtained as the invariant axis of
$\gamma_2\bigl(\gamma_2'(\tau)\bigr)^{-1}$, together with its
identified image in the fundamental domain, and therefore descends
to the closed crossing geodesic in the quotient geometry.
For the parameters shown here, its length at $\tau=0.1$ is
$L_{\Gamma_N}(0.1)=7.22973$, slightly larger than the untwisted
value $L_{\Gamma_N}(0)=2z=7.22445$.
}

\label{fig:schottky-tau01}

\end{figure*}

\section{III. The Twist Lengthens the Crossing Geodesic}

\paragraph*{Independent parameter range for $\tau$.}
The geometry is also invariant under the shift $\tau\to\tau+m/2$,
accompanied by the exchange $B\leftrightarrow B'$. Combined with the
reflection symmetry $\tau\to-\tau$ of Sec.~II, this gives
\begin{equation}
\tau\longleftrightarrow \frac{m}{2}-\tau ,
\end{equation}
so that $\tau=m/4$ is a symmetric point and it is sufficient to
consider the range
\begin{equation}
0\leq\tau\leq\frac{m}{4}.
\end{equation}

It is important, however, that this periodicity of the geometry
should not be confused with periodicity of a fixed crossing curve.
The curve $\Gamma_N$ realizing $AB|A'B'$ tracks a particular
winding class around $M$, and its length $L_{\Gamma_N}(\tau)$ in
Eq.~\eqref{eq:Lcross} increases monotonically with $\tau$ rather
than returning to $2z$ at $\tau=m/2$. 
At $\tau=m/4$, the shortest representatives of the two crossing
partitions $AB|A'B'$ and $AB'|A'B$ become degenerate, as required
by the $B\leftrightarrow B'$ symmetry, which gives
$L^{\rm conn}_{AB}(\tau)=L^{\rm conn}_{AB'}(m/2-\tau)$. As we show
below, $L^{\rm conn}_{AB'}(\tau)\geq L_{\Gamma_N}(\tau)$ for
$0\leq\tau\leq m/4$, so it is sufficient to follow the
$AB|A'B'$ branch, Eq.~\eqref{eq:Lcross}, over this range.

\paragraph*{The actual crossing-geodesic length.}
Paper~I bounded the length of any simple closed curve $\Gamma$
realizing the crossing partition $AB|A'B'$ by $L(\Gamma)\geq2z$, with
$z=z(\ell,m)$ the ($\tau$-independent) length of the shortest
returning orthogeodesic on each side. This bound is a statement
about the shortest arc on \emph{each side separately}; it does not
by itself determine the length of the shortest \emph{closed} curve
obtained by joining the two sides across $M$. At $\tau=0$, the two
shortest arcs join without a kink, and the bound is saturated
exactly by the closed crossing geodesic, $L_{\Gamma_N}(0)=2z$. For $\tau\neq0$, this is no longer automatic: the
twist of the $\{B,B'\}$ pants by $\tau$ shifts the two endpoints on $M$
out of alignment, so that closing the curve requires deforming away
from the individually shortest arcs, and $L_{\Gamma_N}(\tau)>2z$.

This length is computed directly from the trace of the corresponding
group element. The crossing curve realizing $AB|A'B'$ is the closed
geodesic associated with \footnote{The inverse orientation,
$\gamma_2^{-1}\gamma_2'(\tau)$, gives the same trace, since
$\operatorname{Tr}(X^{-1})=\operatorname{Tr}(X)$ for $X\in SL(2,\mathbb R)$.}
\begin{equation}
N(\tau)=\gamma_2\,\gamma_2'(\tau)^{-1}.
\end{equation}
Using the standard trace--length relation, see e.g.~\cite{Maxfield:2014kra},
\begin{equation}
\left|\operatorname{Tr}N(\tau)\right|
=
2\cosh\frac{L_{\Gamma_N}(\tau)}{2},
\end{equation}
a direct evaluation gives
\begin{equation}
L_{\Gamma_N}(\tau)
=
2\operatorname{arccosh}\!\left[
\frac{
c_1c_2+\left(c_1c_2 +R_1R_2 \right) \cosh\tau
}{
R_1R_2
}
\right].
\label{eq:Lcross-direct}
\end{equation}
Equivalently, defining $a\equiv\frac{c_1c_2}{R_1R_2}$,
we obtain
\begin{equation}
L_{\Gamma_N}(\tau)
=
2\operatorname{arccosh}\!\big[a+(a+1)\cosh\tau\big].
\label{eq:Lcross}
\end{equation}
Using the trace conditions $L_A=L_{A'}=\ell$, which on the branch
$c_1,c_2>0$ give $c_1=\mu c_2$ and
$c_2=2\sqrt{R_1R_2}\cosh(\ell/2)/(\mu-1)$, we can write this parameter as
\begin{equation}
a
=
\frac{\cosh^2(\ell/2)}{\sinh^2(m/4)}.
\label{eq:acoshlsinhm}
\end{equation}
At $\tau=0$ this reduces
to $L_{\Gamma_N}(0)=2\operatorname{arccosh}(2a+1)$, which one can
check coincides with $2z$ of Eq.~\eqref{eq:zformula} for any
$(\ell,m)$. Equation~\eqref{eq:Lcross} is even in $\tau$, and
within the independent range $0\leq\tau\leq m/4$,
$L_{\Gamma_N}(\tau)$ increases monotonically away from $\tau=0$,
with $L_{\Gamma_N}(\tau)>2z$ for $\tau>0$.

\paragraph*{$\Gamma_N$ is the shortest crossing curve.}
We now show analytically that $L_{\Gamma_N}(\tau)$ of
Eq.~\eqref{eq:Lcross} is the length of the shortest simple closed
curve with $i(\Gamma,M)=2$ realizing either crossing partition, for
$0\leq\tau\leq m/4$.

Cutting such a curve along $M$ gives one essential arc in each pair
of pants, separating the two external cuffs on that side. As used in
Paper~I, this arc is unique up to isotopy once its endpoints are
allowed to slide along $M$, and sliding an endpoint once around $M$
changes its lift by $\gamma_1$. The arc is therefore the returning
orthogeodesic of Sec.~II, carried by $\gamma_2$ on the $A$ side and by
$\gamma_2'(\tau)$ on the $B$ side, up to powers of $\gamma_1$ at its
ends. Joining the two arcs, the holonomy of the curve is, up to
conjugation and inversion,
\begin{equation}
\begin{aligned}
&\qquad \quad w=\gamma_2\,\gamma_1^{\,n_1}\,X\,\gamma_1^{\,n_2},
\\
&X\in\{\gamma_2'(\tau)^{-1},\,\gamma_2'(\tau)\},
\quad n_1,n_2\in\mathbb Z .
\label{eq:wordfamily}
\end{aligned}
\end{equation}
This family contains every simple curve with $i(\Gamma,M)=2$; it may
also contain non-simple words, which only enlarge the set over which
we minimize.

Multiplying the $2\times2$ matrices, and using $c_1=\mu c_2$, so
that $c_1c_2/(R_1R_2)=a$, $c_2^{\,2}/(R_1R_2)=a/\mu$ and
$c_1^{\,2}/(R_1R_2)=a\mu$, we find, with $\mu=e^{m/2}$,
\begin{equation}
|\operatorname{Tr}w|
=
2a\cosh\frac{Km}{2}+2(a+1)\cosh\!\Big(\tau+\frac{Dm}{2}\Big),
\label{eq:traceKD}
\end{equation}
where $D=n_1-n_2$, and $K=n_1+n_2$ for $X=\gamma_2'(\tau)^{-1}$,
$K=n_1+n_2-1$ for $X=\gamma_2'(\tau)$. In either case $K,D\in\mathbb Z$. The curve $\Gamma_N$ corresponds to $(K,D)=(0,0)$, with
$|\operatorname{Tr}|/2=a+(a+1)\cosh\tau$.

For $0\leq\tau\leq m/4$ we have $\cosh(Km/2)\geq1$, and
$|\tau+Dm/2|\geq\tau$ for every integer $D$: this is obvious for
$D\geq0$, while for $D\leq-1$ one has
$|\tau+Dm/2|\geq m/2-\tau\geq\tau$. Since $a>0$,
\begin{equation}
|\operatorname{Tr}w|\;\geq\;2a+2(a+1)\cosh\tau
=|\operatorname{Tr}\Gamma_N|,
\end{equation}
with equality only for $(K,D)=(0,0)$, that is $\Gamma_N$, and
additionally for $(K,D)=(0,-1)$ at $\tau=m/4$. The latter is
represented by $\gamma_2\gamma_2'(\tau)\gamma_1\sim
\gamma_1\gamma_2\gamma_2'(\tau)$, whose trace equals that of
$\Gamma_N$ evaluated at $m/2-\tau$, in accordance with the
$B\leftrightarrow B'$ symmetry. Hence, for $0\leq\tau\leq m/4$,
$L^{\rm conn}_{AB}(\tau)=L_{\Gamma_N}(\tau)$ and
$L^{\rm conn}_{AB'}(\tau)\geq L_{\Gamma_N}(\tau)$, with equality at
$\tau=m/4$, so that comparing $L_{\Gamma_N}$ alone with $2\ell$
suffices to determine both $S(AB)$ and $S(AB')$.

Curves with $i(\Gamma,M)\geq4$ contain at least four essential arcs,
each of length at least $z$, and hence have $L\geq4z$; this exceeds
$2\ell$ throughout the region considered here (it is equivalent to
$2z>\ell$; e.g.\ $4z=15.04$ at $\ell=m=4$), so they never compete
with the disconnected pair of horizons.

\section{IV. The Bipartite-Invariant Window Widens with the Twist}

\begin{figure}[t]
\centering
\includegraphics[width=0.6\linewidth]{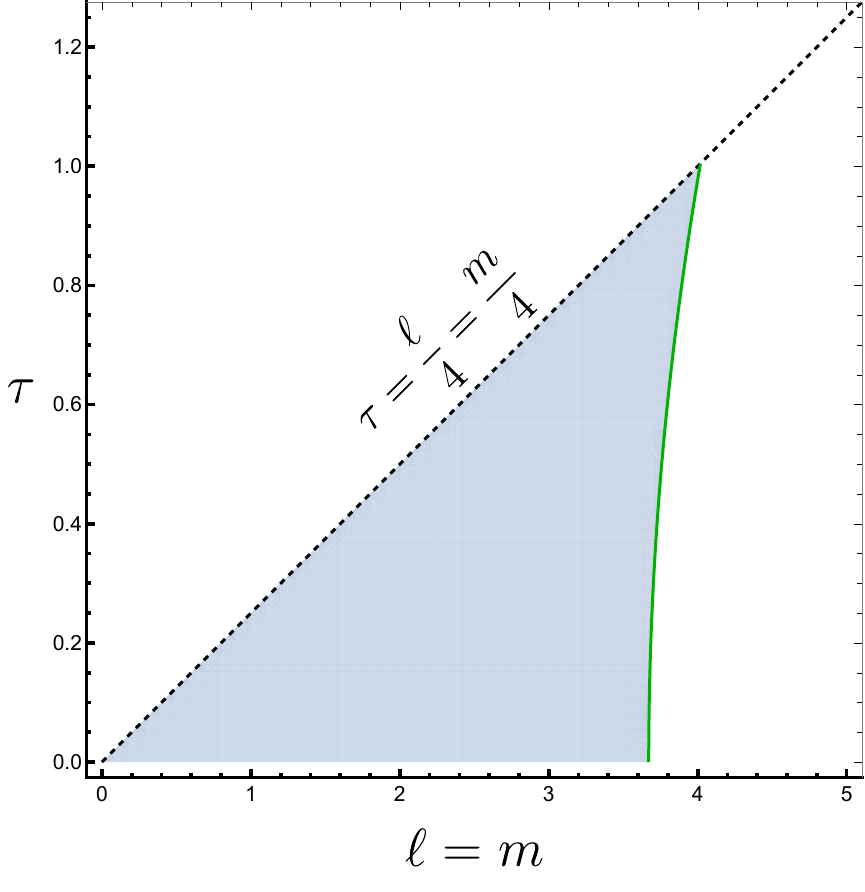}
\caption{The region in the $(\ell=m,\tau)$ plane, restricted to $\tau\in[0,m/4]$, in which the bipartite RT entropies are guaranteed to be exactly $\tau$-independent ($L_{\Gamma_N}(\tau)>2\ell$); the shaded region is where this condition holds, bounded by the green curve. The boundary widens from $\ell_*\simeq3.671274$ at $\tau=0$ to $\ell_c\simeq4.018428$ at $\tau=m/4$.}
\label{fig:window}
\end{figure}

The bipartite RT entropies for $AB,AB'$ remain exactly as in
Eq.~\eqref{eq:SABetc}, $\tau$-independent, whenever the disconnected
candidate wins, $L_{\Gamma_N}(\tau)>2\ell$. On the symmetric line
$\ell=m$, this condition defines a region in the $(\ell,\tau)$ plane,
shown in Fig.~\ref{fig:window}: using
Eq.~\eqref{eq:Lcross}, the boundary of this region is the curve
$2\operatorname{arccosh}[a+(a+1)\cosh\tau]=2\ell$, restricted to the
fundamental domain $\tau\in[0,m/4]$ established in Sec.~III.
Evaluated at the two symmetric points of this domain, the boundary
gives
\begin{align}
(\tau=0):&\qquad 0\leq\ell\leq\ell_*\simeq3.671274,
\label{eq:ellstar}
\\
(\tau=m/4):&\qquad 0\leq\ell\leq\ell_c\simeq4.018428.
\label{eq:ellc}
\end{align}
Thus $\ell_*$, the boundary already identified in Paper~I from the $\tau\to0$ perturbative analysis, is not the actual boundary of the bipartite-invariant window: allowing $\tau$ to range over its full fundamental domain widens the window in $\ell$ from $\ell_*$ to the larger value $\ell_c$, as Fig.~\ref{fig:window} makes explicit. 
In particular, for $\ell_*<\ell<\ell_c$, the connected crossing geodesic wins near $\tau=0$ (in agreement with Paper~I), but as $\tau$ is increased sufficiently close to $m/4$, it becomes longer than the disconnected pair of horizons. The latter then becomes the RT surface, whose length $2\ell$ is exactly $\tau$-independent.
The twist sensitivity of the ${\mathtt q}=4$ network in this window is addressed separately in Secs.~V and~VI.

\begin{figure}[t]
\centering
\includegraphics[width=0.85\linewidth]{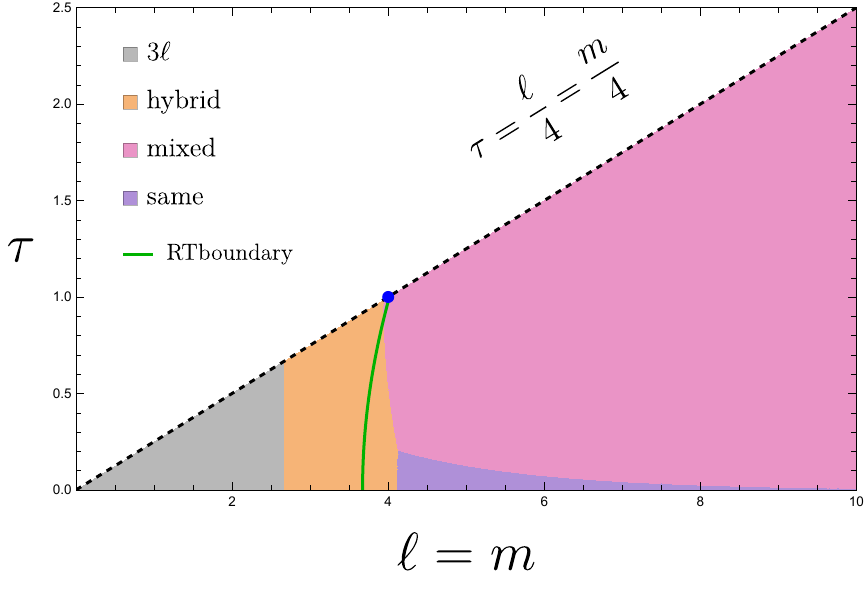}
\vspace{2mm}
\includegraphics[width=0.85\linewidth]{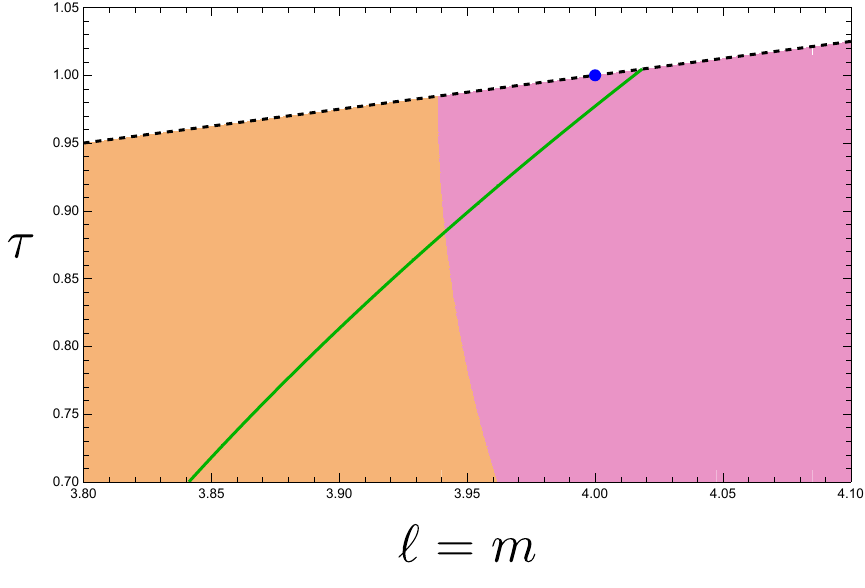}
\caption{Region in the $(\ell=m,\tau)$ plane, restricted to
$0\leq\tau\leq m/4$. To the left of the green curve, the bipartite RT entropies satisfy $L_{\Gamma_N}(\tau)>2\ell$ and are therefore $\tau$-independent, while the twist-sensitive H-tree is the global minimum in the indicated mixed- and same-channel subregions.
Their overlap thus defines a finite parameter region in which $S^{(4)}$ detects the twist although the complete bipartite RT entropy vector does not. The lower plot zooms in on this overlap region. The point $(\ell,\tau)=(4,1)$ (marked by the blue dot) is the benchmark studied in Sec.~V. This figure and Fig.~\ref{fig:ell-m-region} are numerically generated in Mathematica using the analytic methods described in Appendices A and B.}
\label{fig:elltauregion}
\end{figure}

\section{V. A New Benchmark Point: $\ell=m=4$, $\tau=1$}
As a concrete benchmark within the window identified in Sec.~IV, we
examine $\ell=m=4$ in detail; as shown in Sec.~VI, this point lies in
the sub-window in which the mixed-channel network is minimal. Since $m=4$, the symmetric point of
the fundamental domain is at $\tau=m/4=1$; we focus on this point.

\paragraph*{The bipartite side.}
At $\ell=m=4$, the returning-orthogeodesic bound of Paper~I gives
$2z=7.521030\ldots$, already below $2\ell=8$: the sufficient
condition of Eq.~\eqref{eq:crossing-condition} fails at $\tau=0$, in
agreement with $\ell=4$ lying beyond $\ell_*\simeq3.671274$. The
actual crossing-geodesic length of Eq.~\eqref{eq:Lcross}, however,
grows with $\tau$, and at $\tau=1$ we find
\begin{equation}
L_{\Gamma_N}(1)=8.021684\ldots>2\ell=8,
\end{equation}
so that the disconnected pair of horizons wins after all: the
complete bipartite RT entropy vector is exactly $\tau$-independent
at this point,
\begin{equation}
\begin{aligned}
S(A)&=S(A')=S(B)=S(B')=\frac{4}{4G_N},\\
S(AA')&=S(BB')=\frac{4}{4G_N},\\
S(AB)&=S(A'B')=S(AB')=S(A'B)=\frac{8}{4G_N}.
\end{aligned}
\end{equation}
This holds throughout an interval of $\tau$ around $\tau=1$: solving
$L_{\Gamma_N}(\tau)=2\ell$ numerically gives a single crossing at
$\tau_c\simeq0.977$, so that the bipartite entropies are exactly
$\tau$-independent for all $\tau_c<\tau\le m/4=1$ within the
fundamental domain.

Table~\ref{tab:LGammaN} shows $L_{\Gamma_N}(\tau)$ over the full
fundamental domain $\tau\in[0,1]$ at $\ell=m=4$, confirming that it
increases monotonically from $2z=7.521030\ldots$ at $\tau=0$ to
$8.021684\ldots$ at $\tau=m/4=1$. The values obtained from the
closed-form expression, Eq.~\eqref{eq:Lcross}, agree with an
independent direct numerical evaluation of the trace of
$\gamma_2\gamma_2'(\tau)^{-1}$ to all digits shown.

\begin{table}[h]
\centering
\begin{tabular}{c|c|c|c}
\hline\hline
$\tau$ & $L_{\Gamma_N}(\tau)$ & $L_{\rm mixed}(\tau)$ & $L_{\rm same}(\tau)$ \\
\hline
$0.0$ & $7.521030$ & $11.2780$ & $11.1679$ \\
$0.1$ & $7.526265$ & $11.2305$ & $11.1701$ \\
$0.2$ & $7.541943$ & $11.1881$ & $11.1767$ \\
$0.3$ & $7.567973$ & $11.1506$ & $11.1876$ \\
$0.4$ & $7.604209$ & $11.1182$ & $11.2029$ \\
$0.5$ & $7.650453$ & $11.0907$ & $11.2225$ \\
$0.6$ & $7.706456$ & $11.0683$ & $11.2463$ \\
$0.7$ & $7.771925$ & $11.0508$ & $11.2743$ \\
$0.8$ & $7.846530$ & $11.0384$ & $11.3065$ \\
$0.9$ & $7.929912$ & $11.0309$ & $11.3427$ \\
$1.0$ & $8.021684$ & $11.0284$ & $11.3829$ \\
\hline\hline
\end{tabular}
\caption{$L_{\Gamma_N}(\tau)$, $L_{\rm mixed}(\tau)$, and
$L_{\rm same}(\tau)$ at $\ell=m=4$: $L_{\Gamma_N}$ and $L_{\rm same}$
increase monotonically with $\tau$, while $L_{\rm mixed}$ decreases
monotonically. For reference, $L_{\rm hybrid}=11.0839\ldots$ and
$L_{\rm branchless}=12$ are exactly $\tau$-independent.}
\label{tab:LGammaN}
\end{table}


\paragraph*{The $\mathtt q=4$ side.}
We compare the same four competitors as in Paper~I--the branchless
configuration, the hybrid configuration, and the same- and
mixed-channel H-trees--now evaluated at $\ell=m=4$, $\tau=1$.
The H-tree lengths can be evaluated using the direct-edge
representation of Paper~I, Eq.~(54). At the new benchmark point
considered here, we have verified numerically that the front and
back legs cross the corresponding seams transversally, so that this
representation remains valid. For the same-channel and mixed-channel
graphs this gives
%
\begin{align}
L_{\rm same}&=d(v_1,\gamma_2^{-1}v_3)+d(v_1,v_3)
+d(v_2,\gamma_1v_4)\nonumber\\
&\quad+d(v_2,\gamma_2'(\tau)^{-1}v_4) +d(v_1,v_2)+d(v_3,v_4),\\
L_{\rm mixed}&=d(v_1,\gamma_2^{-1}v_3)+d(v_1,v_4)
+d(v_2,\gamma_1v_3)\nonumber\\
&\quad+d(v_2,\gamma_2'(\tau)^{-1}v_4)+d(v_1,v_2)+d(v_3,v_4).
\end{align}
The branchless and hybrid configurations are the same as in
Paper~I and do not depend on $\tau$; at $\ell=m=4$ their lengths are
$L_{\rm branchless}=3\ell=12$ and $L_{\rm hybrid}=11.0839\ldots$. 
The H-tree lengths, by contrast,
must now be extremized directly at $\tau=1$, since the relevant
geometry changes with $\gamma_2'(\tau)$. 
Doing so, within the graph class of Paper~I, Eq.~(54), gives
\begin{equation}
L_{\rm same}(1)=11.3829\ldots,
\qquad
L_{\rm mixed}(1)=11.0284\ldots.
\end{equation}
Collecting these,
\begin{equation}
L_{\rm mixed}<L_{\rm hybrid}<L_{\rm same}<L_{\rm branchless},
\end{equation}
so the mixed-channel H-tree--a configuration that never wins in any
of the comparisons made in Paper~I, where the same-channel graph
always dominates--is the global minimum among the configurations
considered here. Its length depends on $\tau$: $L_{\rm mixed}(0)=
11.2780\ldots$, whereas $L_{\rm mixed}(1)=11.0284\ldots$. Moreover, $L_{\rm mixed}(\tau)$ decreases monotonically from $\tau=0$
to $\tau=m/4$, opposite to the monotonic increase of $L_{\rm
same}(\tau)$ (Table~\ref{tab:LGammaN}), and is smaller than $L_{\rm
hybrid}$ throughout $0.528<\tau\leq m/4$, which contains the whole
interval $\tau_c<\tau\leq m/4$ on which the bipartite entropies are
$\tau$-independent.

\paragraph*{Comment on the $\mathtt q=3$ case.}
For $\mathtt q=3$, there are six possible partitions, which reduce by symmetry to three inequivalent cases. Let us first consider $S^{(3)}(A:B:A'B')$. Although some of these candidate saddles depend on $\tau$, it is possible to draw fairly general conclusions, as shown below. 
We focus on the region where the bipartite entropies are insensitive to the twist $\tau$. Using the known inequality \cite{Iizuka:2025ioc}, any network saddle of $S^{(3)}(A:B:A'B')$ satisfies
\begin{equation}
\begin{aligned}
&S^{(3)}(A:B:A'B')\\
&\quad \geq \frac{1}{2}(S(A) + S(B) + S(A'B')) = \frac{2\ell}{4G_N}.
\end{aligned}
\end{equation}
Here, the right-hand side can be realized not merely as a lower bound, but as a branchless saddle, namely by cutting the horizons
of $A$ and $B$ alone. For other partitions as well, there exists an appropriate branchless saddle that provides a lower bound.
Therefore, whenever the bipartite entropies fail to detect the twist, $S^{(3)}$ is insensitive to it as well and cannot probe the modulus in this setup.
The same argument holds for $\ell\neq m$ and throughout the allowed range of $\tau$. Strictly speaking, for $2\ell > m$, $S^{(3)}(A:A':BB')$ is the only case where there is no single saddle that directly saturates the lower bound. In this case, $4G_N S^{(3)}(A:A':BB')$ is given naively by ${\rm min}\{2\ell, \ell+m, T(\ell,m)\}$ explicitly, where $T(\ell,m)$ denotes the length of the connected component entering the hybrid configuration, such that $L_{\rm hybrid} = T(\ell,m) + \ell$ (See Appendix~A for its detailed definition and computation). Whichever one is chosen, there is no dependence on $\tau$. Therefore, within the setup considered in this paper, where all horizon lengths are equal, $S^{(3)}$ cannot play the same role as $S^{(4)}$ anywhere in the allowed parameter regions.

\paragraph*{Conclusion at this point.}
At $\ell=m=4$, $\tau=1$, all bipartite RT entropies and $S^{(3)}$ are $\tau$-independent, while the minimal $\mathtt q=4$ network is twist dependent. Thus $S^{(4)}$ detects a bulk modulus invisible to bipartite entanglement, realizing the phenomenon sought in Paper~I.

\section{VI. Beyond $\ell=m$}

Nothing in Secs.~II--IV used $\ell=m$: Eqs.~\eqref{eq:Lcross} and
\eqref{eq:acoshlsinhm}, and the minimality proof of
Sec.~III, hold for any $(\ell,m)$. The $\mathtt q=4$ competitors of
Sec.~V are likewise defined for any $(\ell,m)$, with
$L_{\rm branchless}=\min\{3\ell,\,2\ell+m\}$. We work at $\tau=m/4$,
the symmetric point of the fundamental domain, where $L_{\Gamma_N}$ is
largest (and $L_{\rm mixed}$ is smallest at $\ell=m=4$).

\paragraph*{Bipartite side.}
By Eq.~\eqref{eq:Lcross} the bipartite entropies are
$\tau$-independent at $\tau=m/4$ when $L_{\Gamma_N}(m/4)>2\ell$,
which for fixed $\ell$ holds for $m<m_+(\ell)$ (in the range of $m$
considered here). Curves with $i(\Gamma,M)\geq4$ have $L\geq4z>2\ell$
throughout. For large $\ell$, $a\simeq e^{\ell}/(4\sinh^2(m/4))$ gives
\begin{equation}
L_{\Gamma_N}(m/4)-2\ell\;\longrightarrow\;
-2\ln\!\big[2(\cosh(m/4)-1)\big],
\end{equation}
so that
\begin{equation}
\begin{aligned}
m_+(\ell) &\;\longrightarrow\;m_\infty=4\operatorname{arccosh}\tfrac32
=3.849695\ldots \\ &\qquad \qquad \qquad \qquad (\ell\to\infty).
\label{eq:minfinity4arc}
\end{aligned}
\end{equation}

\paragraph*{The $\mathtt q=4$ side.}
Evaluating the same-channel and mixed-channel H-trees, the hybrid and
the branchless configurations with the method of Sec.~V, we find that
the mixed-channel H-tree is the shortest of the four for $m>m_-(\ell)$, in the range of $\ell$ considered (Table~\ref{tab:strip}). And in Appendix C, we show
\begin{equation}
m_-(\ell)\;\longrightarrow\;m_{-,\infty}\simeq 3.8384756854\ldots\quad(\ell\to\infty).
\end{equation} 

\begin{table}[h]
\centering
\begin{tabular}{c|c|c}
\hline\hline
$\ell$ & $m_-(\ell)$ & $m_+(\ell)$ \\
\hline
3.2 & 4.0505 & 4.2606 \\
3.5 & 3.9936 & 4.1436 \\
4.0 & 3.9325 & 4.0217 \\
5.0 & 3.8730 & 3.9110 \\
8.0 & 3.8402 & 3.8527 \\
12.0 & 3.8385 & 3.8497 \\
\hline\hline
\end{tabular}
\caption{At $\tau=m/4$, the mixed-channel H-tree is the shortest
$\mathtt q=4$ network for $m>m_-(\ell)$, while the bipartite
entropies are $\tau$-independent for $m<m_+(\ell)$.}
\label{tab:strip}
\end{table}

\begin{figure}[t]
\centering
\includegraphics[width=1.\linewidth]{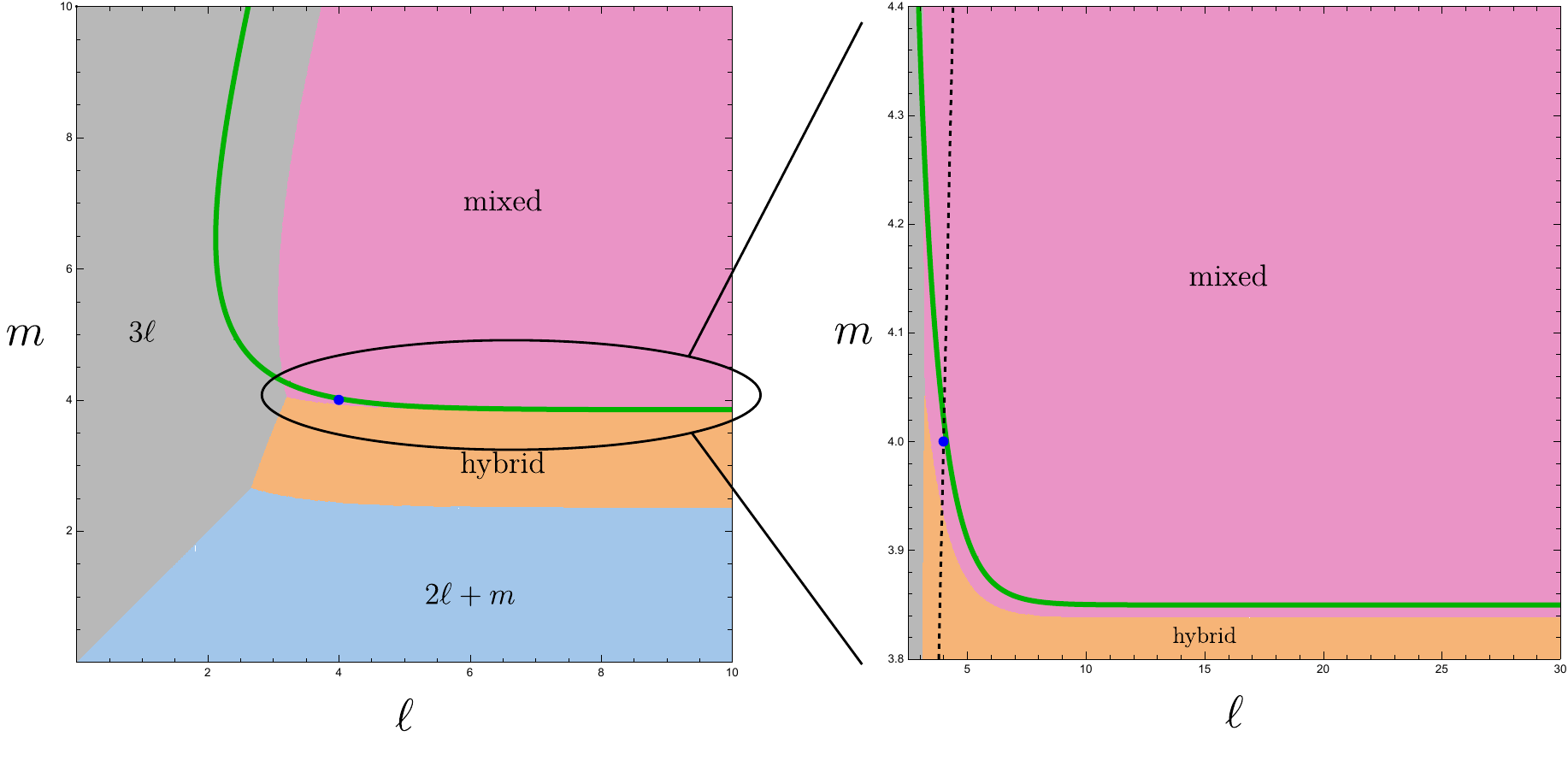}
\caption{Region in the $(\ell,m)$ plane at $\tau=m/4$. The point $(\ell,m)=(4,4)$ is marked by the blue dot. In the region below the green curve, the bipartite RT entropies satisfy $L_{\Gamma_N}(m/4)>2\ell$ and are therefore $\tau$-independent, while the twist-sensitive mixed-channel network is the shortest among the configurations compared here in the indicated mixed subregion. Their overlap thus identifies an open parameter region in which $S^{(4)}$ detects the twist although the complete bipartite RT entropy vector does not. Within this comparison, no same-channel-dominant region appears in the plotted parameter range. The right plot zooms in on this overlap region, and the dashed line denotes $\ell=m$. The overlap persists toward large $\ell$ as a narrow but finite strip in $m$.}
\label{fig:ell-m-region}
\end{figure}

\paragraph*{Result.}
For $\ell\gtrsim3.16$, the interval $m_-(\ell)<m<m_+(\ell)$ is
non-empty: within it the complete bipartite RT entropy vector is
exactly $\tau$-independent while the globally minimal $\mathtt q=4$
network, among the configurations considered, is the twist-sensitive
mixed-channel H-tree. An example is $(\ell,m)=(5,3.90)$, where
$L_{\rm mixed}=12.965$ and $L_{\rm hybrid}=12.988$, and
$L_{\Gamma_N}-2\ell=0.012>0$. The region extends below $\ell_*$, on the side $m>\ell$,
and down to $\ell\simeq3.16$. It narrows as $\ell$ grows: $m_+(\ell)-m_-(\ell)=0.0112$ at $\ell=12$, and the region survives
for $\ell\to\infty$ in a strip $3.838\lesssim m<3.849695$. See Fig.~\ref{fig:ell-m-region}. 
On the line $\ell=m$ it reduces to
$3.9385<\ell<\ell_c\simeq4.0184$.

\section{VII. Summary and Discussion}

In Paper~I~\cite{Iizuka:2026PartI} we found that, near $\tau=0$ and along $\ell=m$, whenever
the complete bipartite RT entropy vector is guaranteed to be
$\tau$-independent, the globally minimal $\mathtt q=4$ network is
$\tau$-independent as well. Here we have shown that this changes at
finite twist. The twist lengthens the crossing geodesic $\Gamma_N$ of
Sec.~III, Eq.~\eqref{eq:Lcross}, and thereby enlarges the region in
which all bipartite RT entropies are exactly $\tau$-independent; near
the symmetric point $\tau=m/4$ this region reaches $\ell_c\simeq
4.018$ on the line $\ell=m$, beyond the $\ell_*\simeq3.671$ of
Paper~I. Within it, the mixed-channel H-tree, which never wins in
Paper~I, becomes the shortest $\mathtt q=4$ network in a sub-window:
at $\ell=m=4$, $\tau=1$ one has $L_{\rm mixed}=11.0284$ against
$L_{\rm hybrid}=11.0839$, $L_{\rm same}=11.3829$ and
$L_{\rm branchless}=12$, while $S(AB)$ and $S(AB')$ are exactly
$\tau$-independent. Away from $\ell=m$ the phenomenon occupies an open
region of the $(\ell,m)$ plane (Sec.~VI), which extends below $\ell_*$,
narrows as $\ell$ grows, and survives as $\ell\to\infty$ in a strip
$3.838\lesssim m<4\operatorname{arccosh}(3/2)\simeq3.8497$.

Two features of the result should be stated plainly. First, the effect
is small. At $\ell=m=4$ the bipartite entropies are $\tau$-independent
for $\tau_c<\tau\leq m/4$ with $\tau_c\simeq0.977$, and on this
interval $L_{\rm mixed}(\tau)$ is smooth and quadratic about $m/4$; it
changes by only $1.3\times10^{-4}$ between $\tau_c$ and $m/4$. The
twist dependence of $S^{(4)}$ is exact, but not large. Second, the margins by which the mixed-channel network beats the hybrid one are modest: $0.055$ at $\ell=m=4$, and typically of order $10^{-2}$ in the large-$\ell$ strip.

This result can be sharpened in terms of genuine multi-entropy.
The genuine four-partite multi-entropy $\mathrm{GM}^{(4)}$ of
Refs.~\cite{Iizuka:2025ioc, Iizuka:2025caq} is obtained from $S^{(4)}$ by subtracting
a linear combination of bipartite and tripartite multi-entropies.
Throughout the bipartite-invariant window, all bipartite entropies are
$\tau$-independent, and Sec.~V shows that all six tripartite
multi-entropies entering $\mathrm{GM}^{(4)}$ are $\tau$-independent
there as well. Consequently, for any two points $\tau_1,\tau_2$
within this window,
\begin{equation}
\mathrm{GM}^{(4)}(\tau_2)-\mathrm{GM}^{(4)}(\tau_1)
=
S^{(4)}(\tau_2)-S^{(4)}(\tau_1).
\end{equation}
This conclusion is independent of the choice of $a$ and $b$ satisfying
$a+b=1/3$ in the definition of $\mathrm{GM}^{(4)}$. Thus the twist
dependence found here is genuinely four-partite: it cannot be
attributed to any bipartite or tripartite contribution.

The conclusion is drawn within the class of configurations
considered: the same- and mixed-channel H-trees in the graph class of
Paper~I, Eq.~(54), the branchless configurations, and the hybrid
configuration, with $L_{\rm branchless}=\min\{3\ell,2\ell+m\}$ for
$\ell\neq m$. For a fixed
graph class the minimization is a convex problem, since
sums of hyperbolic distances are geodesically convex, so
the minima found are global within the class.

The list of graph classes itself is complete, given the
standard fact that optimal Steiner-tree networks are
trivalent (a higher-valence vertex can always be split
into trivalent ones without increasing the total length).
For any admissible network, viewed as a graph on the
compact core with $C$ connected components and $V$
trivalent vertices, Euler's formula $V-E+F=1+C$, together
with $3V=2E$ and $F=4$ (one face per boundary region; here $\mathtt{q}=4$, so there are four boundary regions  $A$, $A'$, $B$, $B'$),
gives $V=6-2C$; since $V\geq0$, only $C=1,2,3$ are possible,
corresponding to $V=4,2,0$ respectively. These are exactly
the mixed/same-channel networks ($C=1$), the hybrid
configuration ($C=2$), and the branchless configuration
($C=3$); no other topology is compatible with a valid
four-region domain partition. Within the $C=1$ class, the
front and back pieces each admit only the two non-crossing
$s$- and $t$-channel topologies of Paper~I, Sec.~IV, which is what
distinguishes the same- and mixed-channel graphs. Thus, given trivalency, the four configurations compared here exhaust the admissible graph topologies that can furnish a minimum, since any reducible network containing redundant bridges can be shortened by deleting those bridges without changing the domain partition. Other windings around $M$ either fail to realize a valid domain partition or are longer, as verified numerically at the benchmark point of Sec.~V.

Several questions remain. Our analysis is restricted to the symmetric family $L_A=L_{A'}=L_B=L_{B'}=\ell$; moving away from it, or to multiboundary wormholes with more asymptotic regions, may enlarge the region. 
And, as noted in Paper~I, the twist need not be invisible to other boundary observables, such as correlation functions between different asymptotic boundaries, which are sensitive to bulk geodesic distances.

\begin{center}
\textbf{Acknowledgments}
\end{center}

\begin{acknowledgments}
T.A.~is supported by JSPS KAKENHI Grant No.~26K17161. N.I.~was supported in part by  NSTC of Taiwan Grant Number 114-2112-M-007-025-MY3, and by MEXT KAKENHI Grant-in-Aid for Transformative Research Areas A “Extreme Universe” No.~21H05184. 
We used Claude (Anthropic, Claude Sonnet 5) and ChatGPT (OpenAI) to assist with the generation and debugging of numerical optimization code, numerical exploration, figure preparation, and language polishing. All AI-assisted code, numerical results, figures, and text were checked and verified by the authors.
\end{acknowledgments}

\begin{center}
\textbf{Appendix}
\end{center}
\section{A. Analytic expression for $L_{\rm hybrid}$}
The hybrid length $L_{\rm hybrid}$ can be written as
\begin{equation}
L_{\rm hybrid}
=\ell+d(v_1,v_2)+d(v_1,\gamma_1v_2)
+d(v_1,\gamma_2^{-1}v_2).
\end{equation}
Here $v_1$ and $v_2$ are the front and back trivalent vertices of the
Steiner tripod on the pair of pants $\{A,A',M\}$, shown in Fig.~4 of
Paper~I (with $v_2$ relabeled as $v_3$ there).
At the minimum, symmetry implies $d(v_1,v_2)=d(v_1,\gamma_1v_2)=r,\ d(v_1,\gamma_2^{-1}v_2)=s$.
For example, at $(\ell,m)=(3.4,4.1)$ we find $r=2.40744$ and $s=1.78089$. To determine $r$ and $s$, it is convenient to express the relevant closed loops in terms of $SL(2,\mathbb R)$ holonomies. 
We define
\begin{equation}
D(d)=
\begin{pmatrix}
e^{d/2}&0\\
0&e^{-d/2}
\end{pmatrix},
\qquad
Q_\theta=
\begin{pmatrix}
\cos\theta&\sin\theta\\
-\sin\theta&\cos\theta
\end{pmatrix}.
\end{equation}
The M\"obius transformation $Q_\theta$ fixes $i$ and rotates its tangent space by an angle $2\theta$.  Since three geodesics meeting at a minimizing trivalent vertex form angles of $2\pi/3$, we set $Q\equiv Q_{\pi/6}$, corresponding to a rotation by $\pi/3$ between successive segments.

Consider first the loop obtained by following
$v_1\to v_2\to\gamma_1^{-1}v_1$.
Viewed globally, this loop is conjugate to the neck holonomy of length
$m$, whereas locally it is represented by two length-$r$ translations
separated by the vertex rotations.  Hence
\begin{equation}
S_v^{-1}(D(r)Q)^2S_v
=
S_n^{-1}D(m)S_n ,
\end{equation}
for some conjugating matrices $S_v$ and $S_n$.  Taking the trace gives
\begin{equation}
\Tr (D(r)Q)^2=2\cosh\frac m2.
\end{equation}
Using $\det(D(r)Q)=1$ and
$\Tr A^2=(\Tr A)^2-2$, we obtain
\begin{equation}
\frac12\Tr[D(r)Q]
=\cosh\frac m4,
\end{equation}
or explicitly
\begin{equation}
\cosh\frac m4
=
\frac{\sqrt3}{2}\cosh\frac r2.
\label{eq:rhybrid}
\end{equation}

Similarly, the loop
$v_1\to v_2\to\gamma_2v_1$ is conjugate to the holonomy of length
$\ell$, while its local representation is
$D(r)QD(s)Q$.  Therefore
\begin{equation}
\frac12\Tr[D(r)QD(s)Q]
=
\cosh\frac\ell2,
\end{equation}
which gives
\begin{equation}
\sinh\frac r2\sinh\frac s2
+\frac12\cosh\frac r2\cosh\frac s2
=
\cosh\frac\ell2.
\label{eq:shybrid}
\end{equation}
Combining Eqs.~\eqref{eq:rhybrid} and \eqref{eq:shybrid}, we find
\begin{align}
&L_{\rm hybrid}=\ell+T(\ell,m),\\
&T(\ell,m) \equiv 2r +s,\\
&r=
2\,\operatorname{arccosh}
\left(
\frac{2\cosh(m/4)}{\sqrt3}
\right),\\
&s=
2\,\operatorname{arcsinh}
\left(
\frac{\cosh(\ell/2)}{\sinh(m/4)}
\right)
-
2\,\operatorname{arcsinh}
\left(
\frac{\cosh(m/4)}
{\sqrt3\,\sinh(m/4)}
\right).
\end{align}
We can use this analytic expression only in the parameter region where $s\geq0$. Also note that this is not an issue of branch selection in the analytic expression, but is already present in \eqref{eq:rhybrid},\eqref{eq:shybrid}. A negative value of $s$ simply indicates that, at those parameter values, the naive hybrid configuration with $120^\circ$ junctions cannot be realized. 
First, $s>0$ always holds for $\ell=m$. For $\ell\neq m$, $s<0$ requires $\cosh(\ell/2)<\cosh(m/4)/\sqrt3$, so in particular $m>4\,\mathrm{arccosh}\sqrt3\simeq4.584863$; since the strip of Fig.~\ref{fig:ell-m-region} lies entirely below $m\simeq4.29$, this never occurs within the strip.

The orientation of the vertex rotation is immaterial for these trace relations. 
Writing $Q_+=Q_{\pi/6}$ and $Q_-=Q_+^{-1}$, one has, for example,
\begin{equation}
\Tr[D(r)Q_+D(s)Q_+]=\Tr[D(r)Q_-D(s)Q_-],
\end{equation}
so either convention gives the same result. 

\section{B. Analytic expression for $L_{\rm mixed}\ \&\ L_{\rm same}$}
The mixed configuration is obtained by minimizing
\begin{align}
L_{\rm mixed} = &d(v_1,v_2)+d(v_3,v_4) +d(v_1,\gamma_2^{-1}v_3) \nonumber\\
&+d(v_1,v_4)+d(v_2,\gamma_1v_3)
+d(v_2,\gamma_2^{\prime-1}v_4)
\end{align}
over $v_1,\ldots,v_4$.
At the symmetric extremum, the six edges pair as
\begin{align}
d(v_1,v_2)&=d(v_3,v_4)\equiv u,\\
d(v_1,\gamma_2^{-1}v_3)
&=d(v_2,\gamma_2^{\prime-1}v_4)\equiv v,\\
d(v_1,v_4)&=d(v_2,\gamma_1v_3)\equiv w,
\end{align}
so that $L_{\rm mixed}=2(u+v+w)$.
For example, at $(\ell,m,\tau)=(3.4,4.1,0.975)$ we find $(u,v,w)=(1.13813,\,2.62608,\,1.19857)$. Unlike the hybrid case, both orientations of the vertex rotation are required.  We define
\begin{equation}
Q_\pm=
\begin{pmatrix}
\frac{\sqrt3}{2}&\pm\frac12\\
\mp\frac12&\frac{\sqrt3}{2}
\end{pmatrix}.
\end{equation}
For a loop composed of two alternating segments of lengths $p$ and $q$,
the relevant holonomy takes the form $H_{pq}=D(p)Q_+D(q)Q_-$. Consider the closed cycle consisting of four segments with alternating lengths $p,q,p,q$, whose geodesic representative has length $W$. Its holonomy is conjugate, up to an overall sign in SL$(2,\mathbb R)$, to $H_{pq}^2$, where $H_{pq}=D(p)Q_+D(q)Q_-$. 
Applying the same trace argument as in Appendix A gives $\frac12|\operatorname{Tr}H_{pq}|=\cosh(W/4)$. It is therefore useful to define
\begin{align}
f(p,q) \equiv \frac12|\Tr H_{pq}| =\frac34\cosh\frac{p+q}{2}
+\frac14\cosh\frac{p-q}{2}.
\label{eq:fpq}
\end{align}
The three independent loops are associated with the neck of length
$m$ and the two nontrivial crossing cycles.  Their geodesic lengths are
\begin{align}
&L_{\Gamma_N}  = 2\,\operatorname{arccosh}
\left[a+(a+1)\cosh\tau\right],\\
&L_{AB'}^{\rm conn} =2\,\operatorname{arccosh}\left[ a+(a+1)\cosh\left(\tau-\frac m2\right) \right],
\end{align}
where $a=\frac{\cosh^2(\ell/2)}{\sinh^2(m/4)}$. Equivalently,
\begin{align}
&\cosh\frac{L_{\Gamma_N}}{4}=\sqrt{1+a}\cosh\frac{\tau}{2},\\ 
&\cosh\frac{L_{AB'}^{\rm conn}}{4}=\sqrt{1+a}\cosh\frac{\tau-m/2}{2}.
\end{align}
Hence $u,v,w$ are determined by solving
\begin{align}
f(u,w)&=\cosh\frac m4,\ \ f(u,v)=\sqrt{1+a}\cosh\frac{\tau}{2},\nonumber \\
f(v,w)&=\sqrt{1+a}\cosh\frac{\tau-m/2}{2},
\label{appb:constraint1}
\end{align}
and the mixed length is then $L_{\rm mixed}=2(u+v+w)$.\\

The same-channel configuration is obtained by minimizing
\begin{align}
L_{\rm same}= &d(v_1,v_2)+d(v_3,v_4)+d(v_1,\gamma_2^{-1}v_3)\nonumber \\
&+d(v_1,v_3) +d(v_2,\gamma_1v_4)
+d(v_2,\gamma_2^{\prime-1}v_4)
\end{align}
over $v_1,\ldots,v_4$. At the extremum, we parameterize the edge lengths by
$d(v_1,v_2)\equiv p,\ d(v_3,v_4)\equiv q,\ d(v_1,\gamma_2^{-1}v_3)=d(v_2,\gamma_2^{\prime-1}v_4)\equiv \alpha,\ d(v_1,v_3)=d(v_2,\gamma_1v_4)\equiv \beta$, so that $L_{\rm same}=p+q+2\alpha +2\beta$. For example, at $(\ell,m,\tau)=(3.4,4.1,0.975)$, we find $(p,q,\alpha,\beta)=(1.53063,\,0.303599,\,2.78489,\,1.43442)$.

To impose the four cycle constraints, define
\begin{equation}
T(x_1^{\sigma_1},\ldots,x_n^{\sigma_n}) \equiv
\frac12\left|
\Tr\prod_{i=1}^n D(x_i)Q_{\sigma_i}
\right|,
\qquad
\sigma_i=\pm.
\end{equation}
The required trace conditions are
\begin{align}
&T(\alpha^+,\beta^+)=\cosh\frac{\ell}{2},\\ 
&T(\beta^+,q^+,\alpha^+,p^+)=\cosh\frac{\ell}{2},\\
&T(\beta^+,q^-,\beta^-,p^+) =\cosh\frac{m}{2},\\ 
&T(\alpha^-,q^+,\alpha^+,p^-)=\cosh\frac{L_{\Gamma_N}}{2}.
\end{align}
Solving these four equations determines $p,q,\alpha,\beta$, and hence $L_{\rm same}=p+q+2\alpha+2\beta$.

\raggedbottom
\section{C. Asymptotic lower boundary}
\label{app:asymptotic-boundary}
We determine the large-$\ell$ limit of the lower boundary $m_-(\ell)$ from the crossing between the mixed and hybrid configurations at $\tau=m/4$. For convenience, define $c=\cosh\frac{m}{4}, h=\sinh \frac{m}{4}$.
At $\tau=m/4$, the two crossing constraints in \eqref{appb:constraint1}  have equal right-hand sides. On the positive-edge branch, they imply $u=w$. Solving the remaining constraints gives
\begin{align}
u&=w=\operatorname{arccosh}\frac{4c-1}{3},\\
v&=2\operatorname{arcsinh}\sqrt{a}-2\operatorname{arcsinh}
\frac{\tanh(m/8)}{\sqrt{3}}.
\end{align}
The mixed length is therefore $L_{\mathrm{mixed}}=4u+2v$, whereas the hybrid configuration has length
\begin{equation}
\begin{aligned}
L_{\mathrm{hybrid}}
={}&\ell+4\operatorname{arccosh}\frac{2c}{\sqrt{3}}
+2\operatorname{arcsinh}\sqrt{a}\\
&-2\operatorname{arcsinh}\frac{c}{\sqrt{3}h}.
\end{aligned}
\end{equation}

For fixed $m$ and $\ell\to\infty$,
\begin{equation}
2\operatorname{arcsinh}\sqrt{a}=\ell-2\ln h+O(e^{-\ell}).
\end{equation}
Thus the terms growing with $\ell$ cancel in the difference,
leaving
\begin{equation}
\begin{aligned}
&\Delta_\infty(m) \equiv\lim_{\ell\to\infty} \bigl[L_{\mathrm{mixed}}(\ell,m,m/4) -L_{\mathrm{hybrid}}(\ell,m)\bigr]\\
&=4\operatorname{arccosh}\frac{4c-1}{3} -4\operatorname{arccosh}\frac{2c}{\sqrt{3}} -2\ln h\\
&\quad-4\operatorname{arcsinh}
\frac{\tanh(m/8)}{\sqrt{3}} +2\operatorname{arcsinh}\frac{c}{\sqrt{3} h}.
\end{aligned}
\label{eq:delta-infty}
\end{equation}
The asymptotic lower boundary is therefore determined by $\Delta_\infty(m_{-,\infty})=0$. Solving this one-variable equation gives
\begin{equation}
m_{-,\infty}
\equiv\lim_{\ell\to\infty}m_-(\ell)
=3.8384756854\ldots,
\end{equation}
with the mixed configuration shorter than the hybrid one
for $m>m_{-,\infty}$. Combining this result with the RT upper boundary $m_\infty=4\operatorname{arccosh}(3/2)$ in Eq.~\eqref{eq:minfinity4arc}, we obtain the nonzero limiting width
\begin{equation}
\lim_{\ell\to\infty}
\bigl[m_+(\ell)-m_-(\ell)\bigr]
=m_\infty-m_{-,\infty}
=0.0112189150\ldots.
\end{equation}

\bibliographystyle{apsrev4-2}
\bibliography{reference.bib}

\end{document}